# A SUPERSTATISTICAL MODEL OF METASTASIS AND CANCER SURVIVAL

L. Leon Chen
*Complex Biosystems Modeling Laboratory, Harvard-MIT (HST) Athinoula A. Martinos Center for Biomedical Imaging, Massachusetts General Hospital, Charlestown, MA 02129, USA*
and
Christian Beck
*School of Mathematical Sciences, Queen Mary, University of London, London E1 4NS, UK*

**ABSTRACT**
We introduce a superstatistical model for the progression statistics of malignant cancer cells. The metastatic cascade is modeled as a complex nonequilibrium system with several macroscopic pathways and inverse-chi-square distributed parameters of the underlying Poisson processes. The predictions of the model are in excellent agreement with observed survival time probability distributions of breast cancer patients.

**Introduction**
Although cancer is a remarkably diverse class of diseases at the genetic level, the similarities become more apparent as we move up the hierarchy of organization. At the highest somatic level, invasion of normal tissue, either adjacent or distant, is essentially the singular means by which cancer realizes its malevolent agenda. The degree to which it is able to carry this out is a function of the states of all the systems lesser in scope. For example, the ability for a cancerous cell to invade is dependent upon certain phenotypic characteristics such as its motile and migratory ability, which is in turn dependent on characteristics such as ability to degrade the extracellular matrix, in turn dependent on relative expression levels of molecules such as matrix metalloproteinases, which is ultimately determined by aberrations in the genetic code. Furthermore, complex epigenetic, evolutionary, and microenvironmental forces act at various levels to help create a complex, highly nonlinear spatially extended dynamical system.

Metastasis is the formation of secondary tumors derived from cells from a malignant primary tumor, and is deadly despite being a highly inefficient process. From a clinical perspective, this inefficiency also makes it an elusive process, as millions of cells are required to disseminate to allow for the selection of cells aggressive enough to survive the metastatic cascade. This cascade is a series of sequential steps, which include the shedding of cells directly into the circulatory system or indirectly via the lymphatic system, survival within the circulation to arrest in a new organ site, extravasation into the new surrounding tissue to initiate growth, and finally induction of angiogenesis to maintain that growth. For a single cell in the primary tumor, metastasis is inherently a stochastic process. This process is driven by an underlying phenotype, resulting in statistical distributions characterized by that phenotype. This distribution starts out being uniformly zero for the case of a cell existing within *in situ* disease. As the accumulation of genetic mutations lead to the acquisition of invasion and migratory mechanisms, this distribution becomes non-zero, and evolves temporally as the acquisition and strengthening of additional mechanisms proceed. Because the stromal microenvironment imposes a selective pressure as well, this distribution can be expected to evolve spatially with respect to cell-microenvironment interaction dynamics.

There has been increasing interest in mathematical modeling of the processes of cancer progression at all different levels, from the theoretical aspects of somatic evolution and its role in carcinogenesis [1-3], to the experimentally validated models of tumor growth and invasion [4-7], to the characteristics of metastatic disease [8-11]. Here, we are concerned with modeling cancer survival, which, as will be explained later, comes out of modeling metastasis.

The model presented here places cancer in the context of a complex nonequilibrium system, and it will be discussed how such a model may generate some insight into the mechanisms of metastasis. Additionally, its potential practical applications in survival analysis are briefly mentioned. Our model is based on superstastical techniques, which are quite frequently used in nonequilibrium statistical mechanics to model complex systems with spatial inhomogenities [19-21]. To the best of our knowledge, the application described here is the first application of superstatistics in medicine, and it is also the first application of a particular class of superstatistics, so-called inverse-chi-square superstatistics.

**Data**
The Surveillance, Epidemiology, and End Results (SEER) database of the US National Cancer Institute was used [12, 13]. Only breast cancer is considered here, which constitutes 717,810 patients diagnosed from 1973 – 2003. Relevant information such as survival time, cause of death, tumor size, histological type, etc. were extracted. Follow-up and accuracy of information is not 100%, but the advantage is clearly its sheer size.

The challenges of developing an accurate theoretical model from experimental analysis are exacerbated by the incredible complexity of the biological system being modeled. Indeed, this is true for any biological system, but it is especially true for cancer systems, due to their multiple levels of organization, heterogeneity from patient to patient as well as within the same patient, and of the number of processes involved. For modeling metastases especially, one needs a large experimental dataset to reduce fluctuations to a minimum, and to allow for the characterization of truly macroscopic behaviors. Therefore, the SEER database is an invaluable resource in this regard.

All patients that are marked as having survival time less than 6 months are excluded from analysis, because the survival time for these patients may not accurately reflect the assumptions of our model. The model is based upon the progression of the metastatic cascade, having undergone treatment in the form of surgery, chemotherapy, or radiation. Although treatment is not explicitly addressed, it is assumed, and acts as another source of heterogeneity in the model. The arbitrary limit of 6 months significantly diminishes the population for whom treatment was delayed, and those who presented with metastatic disease at the time of diagnosis. For patients recorded as having a survival time of zero, results were most likely determined at autopsy, and it is thus not necessarily the case that they died within one month of diagnosis (the resolution of the data).

**The Model**
In over 90% of cancer patients, it is the proliferation of distant metastases that ultimately induce death, and not the primary tumor [14]. Of the remaining 10%, most, if not all, presumably die from some form of recurrence, or failure of primary treatment. Therefore, death can be thought of as the end result of a multiple events of spread, each event characterized by multiple branches, where each branch can be theoretically traced to a progenitor cell in the primary tumor. These progenitor cells can be characterized by the time progressed until it, or any cell the progenitor cells give rise to, becomes a participant in the final events of death. Intuitively, these branches may be thought of as steps in the sequence of events characterizing the progression of cancer leading to death. However, the premise here is that these branches can be eliminated from the model by characterizing only the progenitor cells.

Distant metastases may arise through one of two pathways – through the lymphatic system or the vascular system [15]. There is extensive data indicating that hematagenous metastases and lymphatic metastases obey different statistics [16]. Cancer cells can disseminate from the primary tumor to local lymph nodes, proliferating into micrometastases in these nodes, and then subsequently spreading to distant sites. Alternatively, cancer cells can simply intravasate directly into the bloodstream and end up at distant sites, some surviving to form metastases. Because these two models are complementary to one another, it is assumed that both are occurring to give rise to distant metastases. The extent of the utilization of either pathway varies among cancer types. For example, in the case of breast carcinoma, lymphatic invasion is thought to be the most significant pathway in mid-stage disease, due to the extensive presence of lymphatics [17]. It is thought that the lymph node environment then further selects for more aggressive phenotypes, and the cancer more easily invades distant organs. Of course, in breast cancer as well as other cancers, these pathways are by no means sequential or linear. The simplified process is shown in Figure 1.

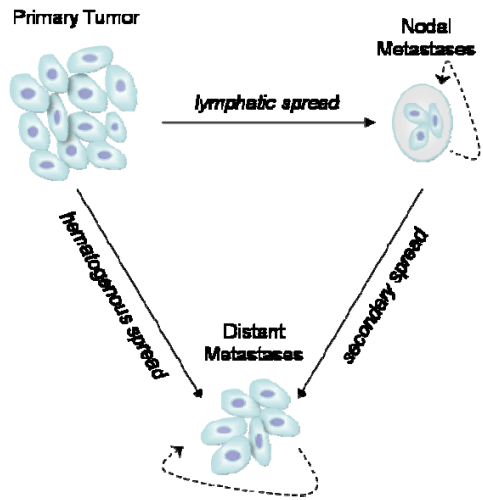

Figure 1. Macroscopic pathways of the metastatic cascade.

Let $T$ be the survival time of a patient. From the definition above, it follows that $T$ is the waiting time until $n$ events of spread have given rise to the collection of metastases present upon death. Each event of spread is an exponentially distributed random variable, assuming it follows a Poisson process. For $n$ exponentially distributed random variables with rate $\lambda$, $T$ follows the gamma distribution [18]:

$$f(t \mid \lambda) = \frac{t^{n-1} \lambda^n \exp(-\lambda t)}{\Gamma(n)}, \tag{1}$$

where qualitatively, $n \geq 1$ controls the shape and $\lambda \geq 0$ adjusts the scale.

Because we are interested in survival, we consider spread events in the macroscopic sense – those that are the precursors to death. There can be any number of spread events, but due to both physical and biological correlations for each pathway of spread, the number of events $n$ can be effectively reduced to an effective dimension of the network of pathways of spread. $n$ thus represents the effective degrees of freedom in Eq. (1). These degrees of freedom are represented by the pathways represented by the solid lines in Figure 1.

However, there are also higher-order pathways in this basic model. For example, cancer cells can spread from one node to other nodes, and cells within distant metastases can even detach and form new metastases [15]. These pathways are shown with the dashed arrows. Fractal properties then emerge within the pathway network, so the effective degrees of freedom for this spread network need not be an integer, and in fact should be greater than 3. However, if a patient's cancer is at a more advanced stage, the effective degrees of freedom should decrease, as the pathways for spread essentially "close off". For example, if a patient presents with nodal metastases, then the pathway from the primary site to the lymph nodes becomes less significant. Random networks in the superstatistical context have also been discussed in [38].

$\lambda$ is an intensive parameter of the cancer system, measuring its intrinsic aggressiveness or invasiveness. However, as a complex non-equilibrium system, dynamical behavior at various microscopic levels will cause spatio-temporal fluctuations in this parameter. It is important to note that this is a function of both a cancerous cell itself as well as its interaction with its surrounding environment. The assumption of superstatistical models is that the system can be partitioned such that each partition will be in local equilibrium, following Boltzmann-Gibbs statistics locally [19-21]. Globally, the system can then be described by a superstatistics, which is a superpositioning of the local statistics of these various partitions each with its own characteristic intensive parameter. Here we generalize this concept to applications in medicine, in particular to the process of metastasis.

The fluctuations in the intensive parameter can be characterized by subordinate processes. For a macroscopic spread event (i.e., an event that ultimately contributes to death), let us assume that there are $s$ participating cells. Each cell will perform a diffusion process. For a diffusion process the mean square displacement grows linearly in time t,

$$\langle x^2 \rangle = Dt, \qquad (2)$$

where $D$ is the diffusion constant. Hence, to travel a typical distance $d$, the corresponding typical time duration satisfies

$$d = (Dt)^{1/2}. \qquad (3)$$

For the spread events, we assume that there are $s$ random variables $\xi_i$ that are approximately independent, each corresponding to the square root of the spread time $t^{1/2}$ for the $i^{th}$ cell. This is illustrated in Figure 2.

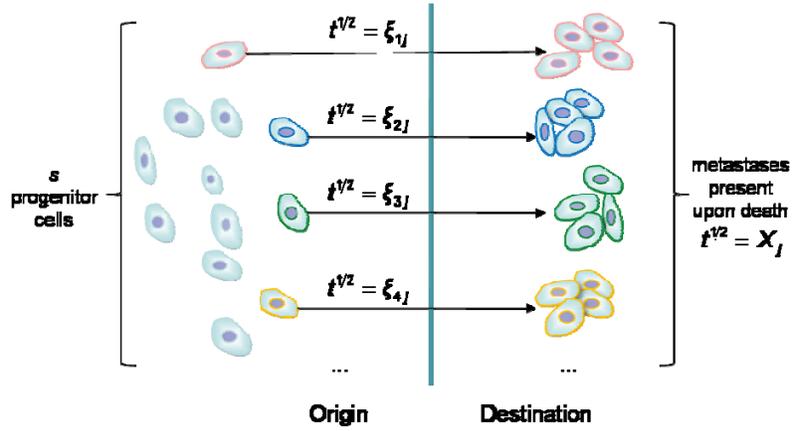

Figure 2. Macroscopic spread events for a degree-of-freedom $j$, which comprises of $s$ participating progenitor cells.

The assumption of independence is valid because the length of time two cells share the same environment is negligibly small. Two cells within the primary tumor can reside within close proximity of one another, and thus their movement may be correlated [22], but as soon as one cell intravasates, they no longer share a single microenvironment. Because a participating cell is defined as one that ultimately contributes to a spread event, it must survive the entire metastatic process, and the time spent at the primary site for these cells is only but a sliver of the metastatic cascade timescale, as it becomes increasingly more difficult to survive within the vasculature and eventually the ectopic host microenvironment after extravasation. It then follows that these random variables are additive, and as such, the sum of these random variables will tend to a Gaussian [21]:

$$\lim_{s \to \infty} \frac{1}{\sqrt{s}} \sum_{i=1}^{s} \xi_{ij} = X_j, \qquad (4)$$

where $X_j$ is a random variable corresponding to the $j^{th}$ degree of freedom in the network of possible paths, $j \in [1, n]$.

Therefore, $1/\lambda$, the expected time to spread, which is the sum of the squared Gaussian random variables $X_j^2$, is $\chi^2$ distributed, and $\lambda$, the rate of spread, is inverse-$\chi^2$ distributed, with degrees of freedom equal to $n$:

$$f(\lambda) = \frac{\lambda_0 (n\lambda_0/2)^{n/2}}{\Gamma(n/2)} \lambda^{-n/2-2} \exp\left(\frac{-n\lambda_0}{2\lambda}\right). \qquad (5)$$

This accounts for fluctuations among different degrees of freedom, or pathways of spread. For example, some patients may present with extensive distant disease without any positive lymph nodes, while others will have micrometastases that remain dormant for extended periods of time [15, 23, 24].

To find the probability density function of survival times, we multiple the expressions in Eqs. (1) and (5) and integrate over all possible $\lambda$:

$$f(t) = \int_0^\infty \frac{t^{n-1}\lambda^n \exp(-\lambda t)}{\Gamma(n)} \frac{\lambda_0 (n\lambda_0/2)^{n/2}}{\Gamma(n/2)} \lambda^{-n/2-2} \exp\left(\frac{-n\lambda_0}{2\lambda}\right) d\lambda, \tag{6}$$

obtaining:

$$f(t) = \frac{(n\lambda_0)^{3n/4}}{\Gamma(n)\Gamma(n/2)} \left(\frac{t}{2}\right)^{3n/4-1} \left[\frac{\sqrt{2n\lambda_0 t}}{n} K_{n/2+1}\left(\sqrt{2n\lambda_0 t}\right) - K_{n/2}\left(\sqrt{2n\lambda_0 t}\right)\right], \tag{7}$$

where $K_\nu(z)$ is the modified Bessel function of the second kind of order $\nu$.

Because the model is of cancer-specific death, this probability density function is conditional upon death from cancer. One can imagine many other possible causes of death, any one of which may occur before the time of death which would have been the result of cancer alone. This is simply due to the stochastic nature of cancer progression, the complex dynamics of comorbidities, and chance events such as accidents. Therefore, this does not model the survival of cancer in the traditional overall sense, but rather finds the time distribution of disease-specific risk.

**Asymptotic behaviors**
It should be noted that the probability density function for $\lambda$ (Eq. (5)) generates the asymptotic form

$$f(\lambda) \sim \exp\left(\frac{-n\lambda_0}{2\lambda}\right) \tag{8}$$

for $\lambda \to 0$. Using Eq. (8), we can then derive the asymptotic behavior of $f(t)$ as $t \to \infty$. The behavior for large times is determined entirely by the behavior for partitions of the system with low rates of metastasis. Rearranging Eq. (6) such that we can apply Laplace's method [25, 26], we have:

$$f(t) = \int_0^\infty \frac{t^{n-1}\lambda^n \exp(-\lambda t)}{\Gamma(n)} f(\lambda) d\lambda = \frac{1}{\Gamma(n)} \int_0^\infty \exp\left[-\lambda t + \ln t^{n-1} + \ln \lambda^n + \ln f(\lambda)\right] d\lambda. \tag{9}$$

The approximation of this integral for large $t$ gives

$$f(t) \sim \frac{\exp\left[-\hat{\lambda}t + \ln t^{n-1} + \ln \hat{\lambda}^n + \ln f(\hat{\lambda})\right]}{\sqrt{-\frac{d^2}{d\hat{\lambda}^2} \ln f(\hat{\lambda})}}, \tag{10}$$

where $\hat{\lambda}$ is the value of $\lambda$ that maximizes the exponential, which equals:

$$\hat{\lambda} = \frac{n + \sqrt{n^2 + 2n\lambda_0 t}}{2t}. \tag{11}$$

Thus, we obtain the following for the numerator and denominator of Eq. (10):

$$\exp\left[-\hat{\lambda}t + \ln t^{n-1} + \ln \hat{\lambda}^n + \ln f(\hat{\lambda})\right] \sim t^{n/2-1} \exp\left(-\sqrt{2n\lambda_0 t}\right), \tag{12}$$

$$\sqrt{-\frac{d^2}{d\hat{\lambda}^2} \ln f(\hat{\lambda})} \sim t^{3/4}. \tag{13}$$

Substituting (12) and (13) back into (10), it is found that as $t \to \infty$, $f(t)$ behaves as:

$$f(t) \sim t^a \exp\left(-b\sqrt{t}\right) \sim \exp\left(-b\sqrt{t}\right) \tag{14}$$

(see also [26]).

We can also determine the asymptotic behavior of the system as a function of its size. Let us define a parameter $\theta = N/\lambda$ where $N$ is the number of cancerous cells. Since the expected time $1/\lambda$ to spread decreases as the number of cells $N$ increases, we may assume that $\theta = N/\lambda$ is constant or nearly constant. In our model, $1/\lambda$ is distributed according to a $\chi^2$ distribution, meaning

$$f(1/\lambda) = \frac{(n\lambda_0/2)^{n/2}}{\Gamma(n/2)} \lambda^{-n/2+1} \exp\left(\frac{-n\lambda_0}{2\lambda}\right). \tag{15}$$

Therefore, as $\lambda \to \infty$, the probability density of $1/\lambda$ decays as a power law.

If the probability of spread as a function of cell number $f(N)$ also behaves as a power law for large N,

$$f(N) \sim N^b, \tag{16}$$

then due to proportionality ($\theta = N/\lambda$ constant) we have

$$b = -n/2 + 1. \tag{17}$$

In a study of the relationship between metastasis probabilities and tumor size [27], it was observed that the per-cell probability of lethal spread of breast cancer and melanoma cells as a function of the number of cells in the tumor displayed a remarkable abidance to a power law. The number of cells for a tumor was extrapolated from its clinical size (roughly the diameter, assuming spherical geometry) at diagnosis, but the fit is unquestionable, with melanoma decaying as $\sim N^{-0.7836}$ and breast cancer as $\sim N^{-0.5611}$, where $N$ is the number of cells. Using Eq. (17), it can then be predicted that $n$ should be on the order of 3.567 for melanoma and 3.122 for breast cancer. In the next section, we see that these values are in agreement with parameters obtained from survival distribution fits.

**Distribution Fits**

Using a simple genetic algorithm method, parameters were obtained for a few breast cancer subpopulations. The objective fitness function is a sum of weighted least squares, with the weights counterbalancing the increasing variance at longer times due to the increasing sparseness of data. First, fitting the model to the entire breast cancer population gives parameter values of $n = 3.00004$ and $\lambda_0 = 0.08712$. The fitted curve as given by Eq. (7) together with the data is plotted in Figure 3. Excellent agreement is obtained.

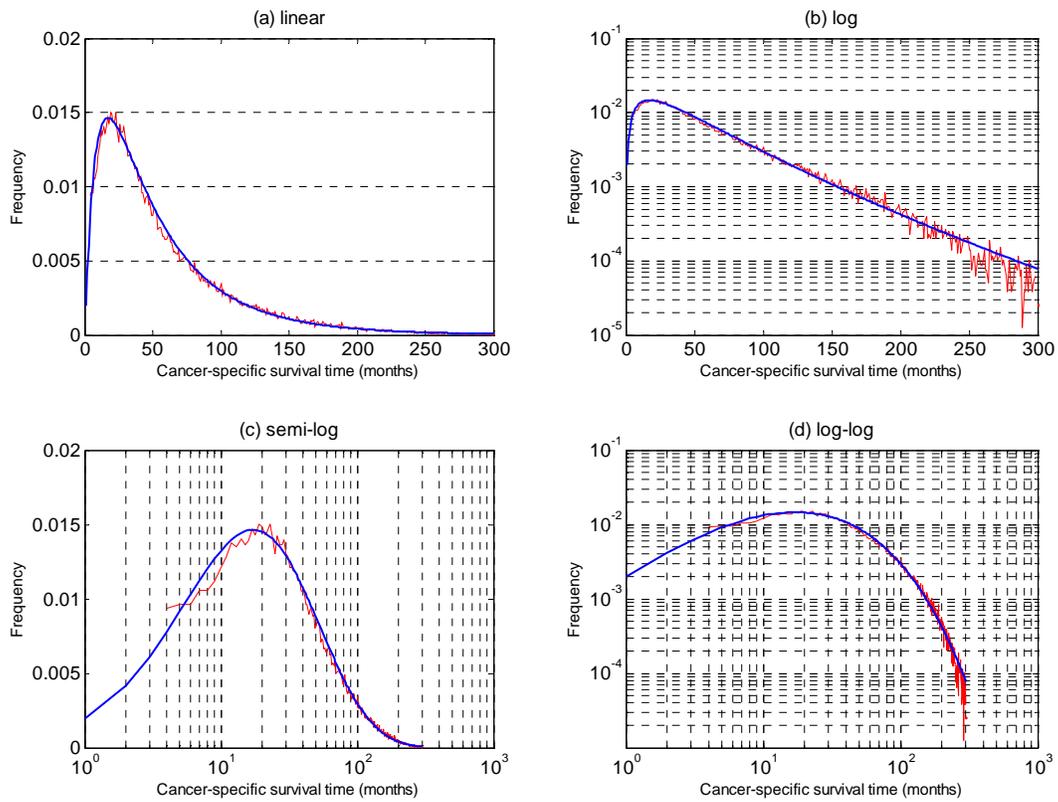

Figure 3. All breast cancer-specific death data (red) and best-fit model (blue): $n = 3.00004$ and $\lambda_0 = 0.08712$.

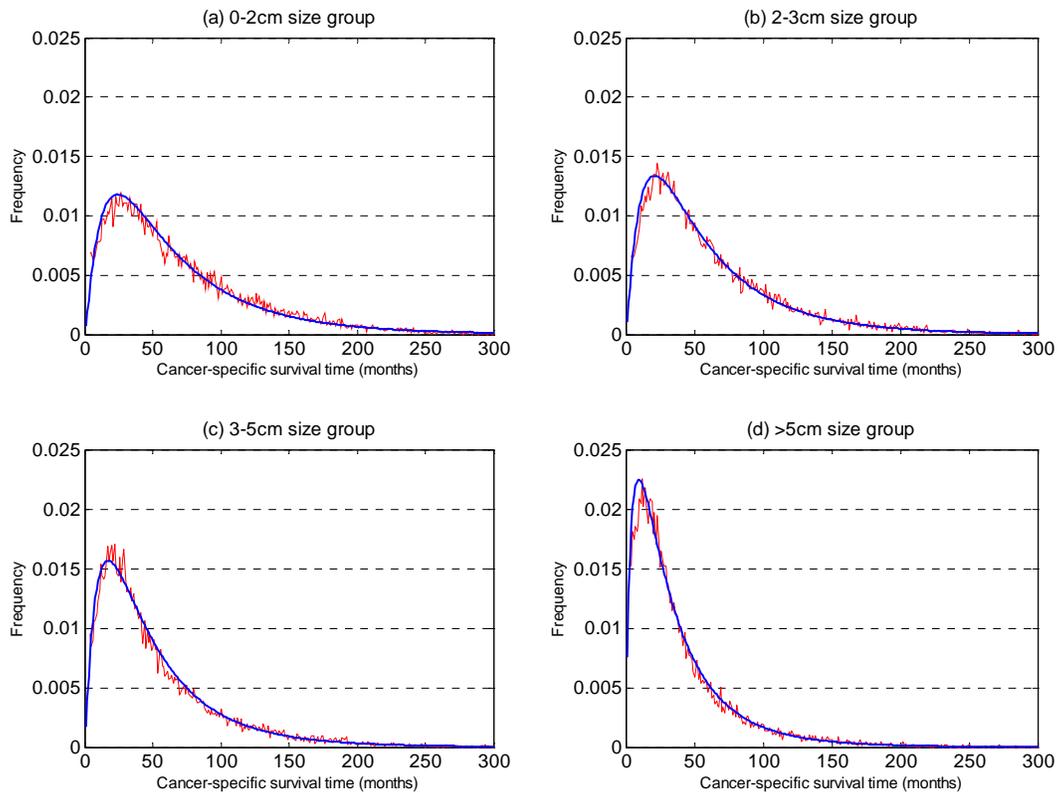

Figure 4. Breast cancer-specific death data divided into comparable-size population groups based on tumor diameter (red) and best-fit models for each (blue).

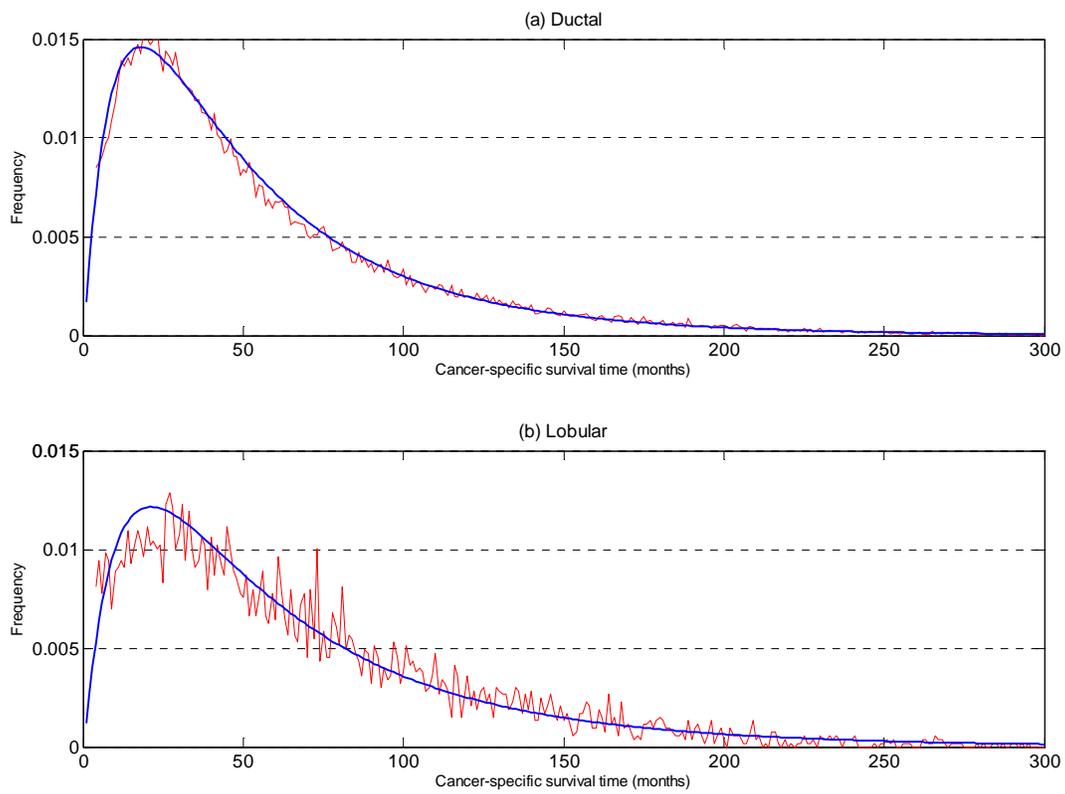

Figure 5. Breast cancer-specific death data of the two most common histological types (red) and best-fit models for each (blue).

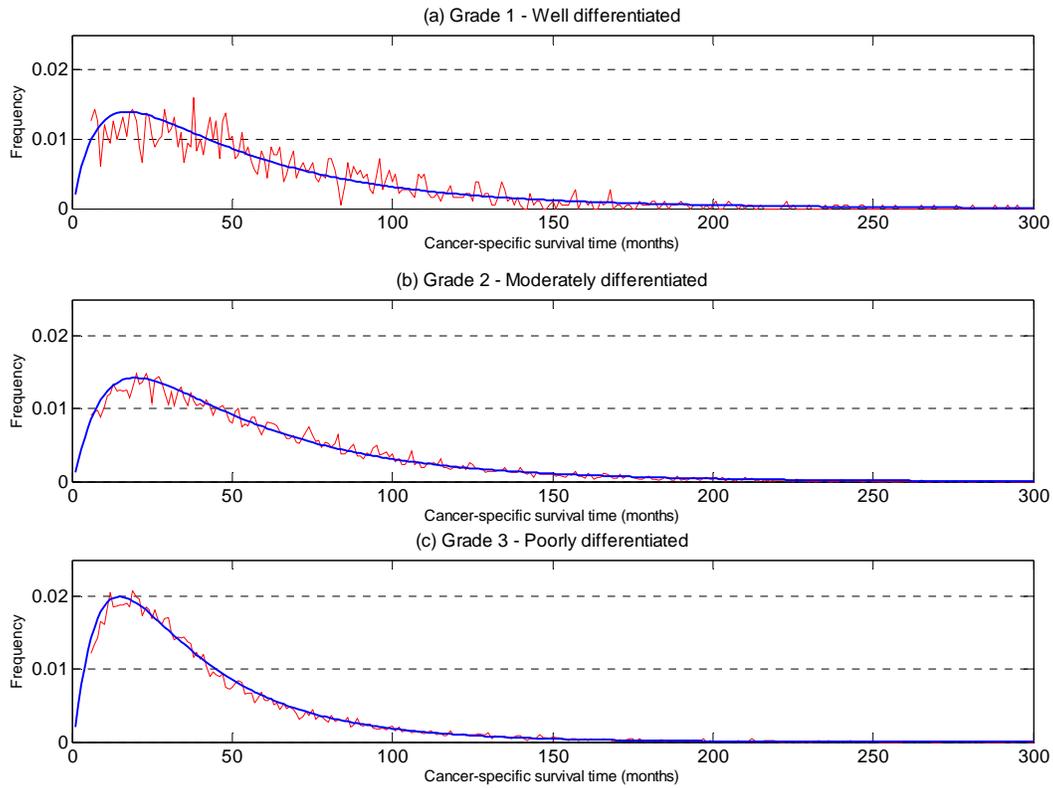

Figure 6. Breast cancer-specific death data divided into different population groups based on pathological grade.

Table 1. Summary of parameters for best-fit superstatistics model of various breast cancer populations.

| Population group | Number of patients | $n$ | $\lambda_0$ | $R^2$ |
|---|---|---|---|---|
| All | 78,779 | 3.00004 | 0.08712 | 0.996 |
| Size: 0-2cm | 19,439 | 3.40346 | 0.07721 | 0.987 |
| Size: 2-3cm | 20,435 | 3.32542 | 0.08605 | 0.989 |
| Size: 3-5cm | 22,004 | 3.21949 | 0.09827 | 0.989 |
| Size: >5cm | 16,793 | 2.58853 | 0.11877 | 0.990 |
| Histology: ductal | 58,398 | 3.13148 | 0.08982 | 0.996 |
| Histology: lobular | 5,287 | 3.07063 | 0.07370 | 0.954 |
| Grade: 1 | 1,815 | 2.89975 | 0.08092 | 0.885 |
| Grade: 2 | 12,586 | 3.34077 | 0.09201 | 0.982 |
| Grade: 3 | 24,544 | 3.51748 | 0.13387 | 0.992 |

Tab. 1 and Figs. 4-7 show the best-fit models for different sub-populations. Because the two parameters in the model correspond to biologically and physically relevant phenomena, it should be expected that they create trends commensurate with the characteristics of the sub-populations. $n$, the effective degrees of freedom in the system, is a more physically-based parameter, while $\lambda_0$, a representation of the aggressiveness of the malignant cells, is a more biologically-based parameter. A larger $\lambda_0$ with *n* held constant corresponds to a higher death rate, as can be expected, but a lower *n* with a constant $\lambda_0$ contributes to a higher death rate as well. The increasing $\lambda_0$ with increasing size and grade is consistent with the correlation between these two prognostic factors and mortality [28]. It is interesting to observe opposite trends for size and grade with respect to *n*, however. One possible explanation may be that with increasing pathological grade, cells become less differentiated, and consequently increase the effective degrees of freedom in the system due to less specific interactions.

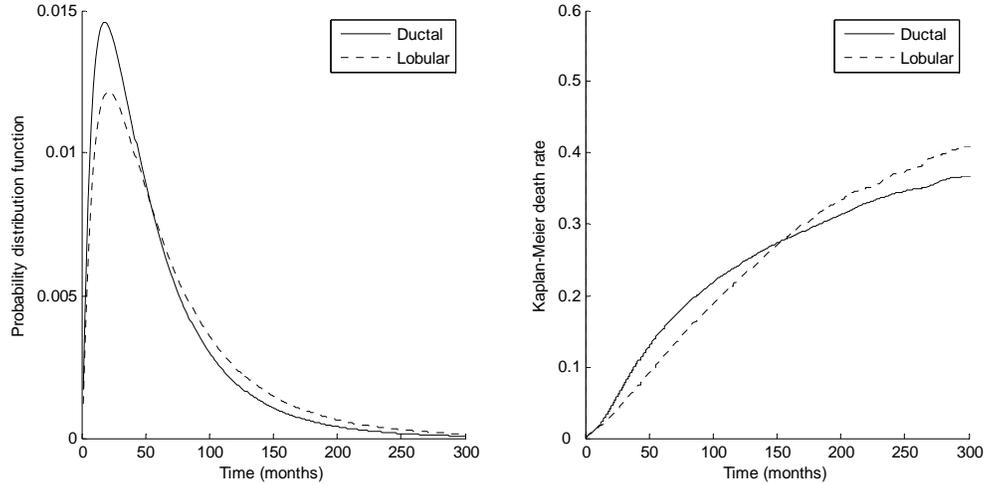

Figure 7. The probability distribution functions using the parameters found in Table 1 are plotted on the left for two histological types: ductal and lobular. The corresponding Kaplan-Meier death rate curves are shown on the right for comparsion. This illustrates the influence of the two parameters on the shape of the resulting distribution. Ductal has a higher $\lambda_0$ value which increases its death rate early on, but lobular has a lower *n*, which causes it to catch up and exceed the death rate for ductal.

**Discussion**

Superstatistics allows one to powerfully characterize fat tails at long times and distances, a hallmark of complex non-equilibrium systems. In cancer-specific survival data, the tail of the frequency distribution decays as an exponential of the square root of time. It has been shown that power-law decays naturally emerge when the fluctuating intensive parameter is $\chi^2$ distributed [19], whereas our data here indicate an inverse $\chi^2$ distributed parameter. Of course, other possibilities exist as well, for example when microscopic contributions to an intensive parameter are multiplicative, in which case the intensive parameter will be lognormally distributed [19-21]. We have demonstrated here for the first time the existence of a complex system that is well described by inverse-$\chi^2$ superstatistics.

The derivation of a dimension parameter for the network of spread pathways was based on a physical argument. It may be plausible to invoke a biological argument, saying that the network of spread pathways is instead the result of different classes of biochemical mediators. For example, factors involved in the induction of angiogenesis and growth factors involved in proliferation may create two distinct 'pathways' a cell may interact upon on its way to become a distant metastasis. Regardless, either formulation of this network leads to an effective dimensional parameter due to correlations, but how it arises will require actual experimental studies.

This paper has attempted to characterize the dynamical reasons for the emergence of the observed cancer disease-specific mortality probability distributions. One interesting hypothesis for how the distribution comes about is the hypothesis that surgical treatment induces wound-healing responses that removes inhibitors of angiogenesis and stimulates production of mitogenic growth factors [29, 30]. This will in turn accelerate the development of dormant distant micrometastases, and lead to a peak hazard a couple of years after surgical treatment. Although it could be a contributor to the statistics of the system, in effect reducing the expected times to spread, there can be a number of other contributors as well. The overall statistics of the cancer system is likely to be the result of the complex interaction between biological factors and physical factors that influence the dynamics of the underlying mechanisms, rather than any single factor alone. For other models dealing with survival-time statistics of cancer patients, see [31-37].

The analysis presented here is based on only the breast cancer data subset, but the extendibility of the model to other cancer subtypes will be a topic for future studies. There have already been hints to the existence of a 'universal' descriptor for cancer growth [7], and its survival distributions [37]. Computational modeling of cancer as a complex non-equilibrium system will help us gain additional insights into its underlying mechanisms, which allows us to better predict outcomes, ultimately leading to incrementally better therapeutic options for the cancer patient.

## ACKNOWLEDGEMENTS

L.L.C. would like to thank James Michaelson and Yehoda Martei for making the publicly-available SEER dataset into a more user-friendly format, and for helpful discussions.